\documentstyle[12pt]{article}
\normalbaselines
\begin{document}
\bibliographystyle{unsrt}
\vbox {\vspace{6mm}}
\begin{center}
{\bf Experimental limit on the blue shift of the frequency of light
implied by a q-nonlinearity}
\end{center}

\begin{center}
V. I. Man'ko\footnote{On leave from Lebedev Physical
Institute, Moscow, Russia.} and G. M. Tino
\end{center}

\bigskip

\begin{center}

Dipartimento di Scienze Fisiche, Universit\'a di Napoli "Federico II"\\
Mostra d'Oltremare, Pad.20 - 80125 Napoli, Italy
\end{center}

\begin{abstract}
We discuss the implications of an experiment in which
the frequencies of two laser beams are compared for different
intensities in order to search for a dependence of the frequency
of light on its intensity. Since
no such dependence was found it is possible to place bounds on
a description of the electromagnetic field in terms of q-oscillators.
We conclude that the value of the nonlinearity parameter is smaller
than $10^{-17}~$.
\end{abstract}

\section{Introduction}
\label{s1}

The q--oscillators \cite{bie}, \cite{mc} have been introduced in the
study of the mathematical structures of quantum groups \cite{jimbo} which
are used to explain the peculiarities of the rotational spectra of
nuclei \cite{smirnov}. The q--deformation changes the
Brownian motion \cite{mendes}. The physical nature of q--oscillators
is not clear yet. It is related sometimes to the problem of fundamental
length \cite{shabanov}, \cite{vitiello} or to the discrete structure of
time \cite{palak}. The quantum group structure may be related to
nonlinear modifications of quantum mechanics \cite{kostin},
\cite{weinberg}, \cite{doebner}, reviewed in \cite{mizrahi}.
The nonlinearity of vibrations produces also changes in the Bose
distribution \cite{su}, \cite{tuszy}.
In \cite{sol1}, \cite{marmo} it was suggested an interpretation of the
q--oscillators as usual nonlinear oscillators with specific exponential
dependence of the oscillator frequency on the amplitude of vibrations. Such
hypothesis leads to the same algebraic structure of the commutation
relations in the quantum case and Poisson brackets in classical domain that
q--oscillators of \cite{bie}, \cite{mc} have. Some predictions have been
formulated on the basis of this hypothesis. If  q--nonlinear vibrations
are considered as vibrations of the real electromagnetic field, they must be
''visible'' for high field intensity as for
nonlinearities of any kind. In particular, a dependence of light frequency
on its intensity has been proposed in \cite{sol2}. It implies that for
increasing electromagnetic field intensity the frequency increases, which
is not the case, of course, for linear Maxwell electrodynamics. While the
nonlinear phenomena in media electrodynamics are well known and they are
the base for nonlinear optics, the q--nonlinearity considered in this work
would be a property of the electromagnetic field vibration in vacuum.

In this paper, we discuss in the frame of the theory of q-nonlinearity
the results of a simple experiment which allows
to set an upper limit to
a possible dependence of the frequency of light on the intensity.
A heterodyne technique is used to compare the frequency of two laser
beams as their intensity is changed. Two narrow-band diode lasers are
phase-locked with a fixed frequency offset and a portion of their output
is mixed in a fast photodiode.
The measurement of the frequency of the beat note provides a very
sensitive method to detect small frequency changes.
Although the work is motivated by the interpretation of q--oscillators as
describing the q--nonlinear vibrations of electromagnetic field, the
accuracy of the experiment allows to set an upper limit to any
dependence of the ''colour'' of light on the intensity.

In the next section, we describe shortly the theory of q--oscillators and
review the results of \cite{sol1}, \cite{marmo} and \cite{sol2}.
In Section 3, the experimental set--up is described. The results of
the experiment are discussed in Section 4.

\section{q--vibrations}
The photon creation and annihilation operators $~a~$ and $~a^{\dag }~$
satisfy boson commutation relations
\begin{equation}\label{v1}
aa^{\dag }-a^{\dag }a=1.
\end{equation}
The one--mode Hamiltonian describing the electromagnetic field
oscillations is taken as the Hamiltonian of the usual harmonic oscillator
\begin{equation}\label{v2}
H=\frac {1}{2}\hbar \omega (a^{\dag }a+aa^{\dag }),
\end{equation}
that yields the Heisenberg equation of motion for the amplitude
$~a~$
\begin{equation}\label{v3}
\dot a(t)=-i\omega a(t).
\end{equation}
The solution of this linear equation is
\begin{equation}\label{v4}
a(t)=a(0)e^{-i\omega t},
\end{equation}
and it is seen from this solution that for linear oscillations the
frequency does not depend on the amplitude of vibrations.

The q--oscillator is described by the creations and annihilations
operators $~a_{q},~a_{q}^{\dag }~$ which satisfy the following
relation \cite{bie}, \cite{mc}
\begin{equation}\label{v5}
a_{q}a_{q}^{\dag }-qa_{q}^{\dag }a_{q}=q^{-n}.
\end{equation}
Here the real number $~q=\exp \lambda ~$, the parameter $~\lambda ~$
is considered as the nonlinearity parameter \cite{sol1}, and the operator
$~\hat n=a^{\dag }a~$ is the operator of the photon number in the mode.
In \cite{sol1}, \cite{marmo} and \cite{sol2} it was suggested using
the nonlinear relations
\begin{equation}\label{v6}
a_{q}=a\sqrt {\frac {\sinh (\lambda a^{\dag }a)}
{a^{\dag }a\sinh \lambda }},~~~~~a_{q}^{\dag }=\sqrt {\frac {\sinh (\lambda a^
{\dag }a)}{a^{\dag }a\sinh \lambda }}a^{\dag },
\end{equation}
which give relation (\ref{v5}). The modified Hamiltonian
(\ref{v2}) has the form
\begin{equation}\label{v7}
H_{q}=\frac {1}{2}\hbar \omega (a_{q}^{\dag }a_{q}+a_{q}a_{q}^{\dag })
=\frac {\hbar \omega }{2\sinh \lambda } \{\sinh [\lambda(a^{\dag }a+1)]
+\sinh (\lambda a^{\dag }a) \}.
\end{equation}
It describes nonlinear vibrations given by the Heisenberg equation of motion
\begin{equation}\label{v8}
a\dot (t)=-i\omega [n]a(t),~~~~~~~a(0)=a.
\end{equation}
The solution to this equation has the form
\begin{equation}\label{v9}
a(t)=a(0)\exp \{-i\widetilde \omega [a^{\dag }(0)a(0)]t\},
\end{equation}
The frequency of vibration $\widetilde \omega$ depends on the
amplitude and for large field intensities $~(n\gg 0)$ it has the form
\begin{equation}\label{v10}
\widetilde \omega =\omega \frac {\lambda }{\sinh \lambda }\cosh \lambda n.
\end{equation}
In the limit of small nonlinearity, $~\lambda \rightarrow 0~$
and the frequency $~\widetilde \omega \rightarrow \omega .$

The nonlinearity parameter $~\lambda ~$ is supposed to be small and the
"blue shift effect" \cite{marmo}\cite{sol2}
which is interpreted as a change of the
frequency due to increasing number of photons in the mode is described
by
\begin{equation}\label{v11}
\frac {\delta \omega }{\omega }\approx \frac{1}{2}\lambda ^{2}n^{2},
{}~~~~~~\delta \omega =\widetilde \omega -\omega,
\end{equation}
which is the approximate expression (\ref{v10}) for $~\lambda \ll 1~$,
$~n\gg 0$ and $~\lambda n \ll 1~$.
This relation allows to check experimentally whether
there is a change of light frequency if one increases the photon number
in the mode (field intensity). A null result in the search for a frequency
shift provides an upper bound for the nonlinearity parameter $~\lambda~$.

\section{Experiment}

The basic idea of this experiment is to search for a change of the frequency of
laser radiation when its intensity is varied. In this experiment, this was
done by comparing the frequencies of two laser beams using a heterodyne scheme.
The output of two narrow-band semiconductor diode lasers
was mixed in a fast photodiode.
By varying the intensity of the beams, a dependence of the frequency of the
beat
note on the intensity was searched for.  A very high accuracy was obtained by
phase-locking one of the lasers to the other.  In this way, the frequency
difference at the output of the two laser sources was extremely stable and
small
changes could be observed.

  A simplified scheme of the apparatus used for the measurements reported in
this paper is shown in Fig. 1.  Two diode lasers emitting at 850
nm were used as radiation sources. In order to reduce their linewidth, they
were
mounted in an extended-cavity configuration using optical feedback from an
external grating \cite{tino}. This allowed to reduce the linewidth
to about 50 kHz. The output of the lasers was the zero-order reflection from
the grating. The output beam of each laser passed through an optical isolator,
to avoid feedback to the laser, and through an anamorphic prism pair to correct
the elliptical cross-section of the beam. At the output of the prisms the beam
had a near-circular cross-section with a diameter of 8 mm.

The two diode lasers were phase-locked with a frequency offset of 9 GHz using
the scheme described in \cite{santarelli}. When two lasers are phase-locked,
the frequency difference becomes equal to the frequency of the reference
signal.
The relative linewidth of the locked lasers is therefore negligible and is only
limited by the linewidth of the frequency generators and by vibrations of
optical components. In the set-up we used for this experiment, under locked
conditions the measured phase error variance was less than $4.10^{-3} rad^{2}.$

The frequency difference between two laser beams, each coming from one laser,
was measured by measuring the frequency of the beat note obtained at the output
of the photodetector (PD2) on which the two beams were superposed. The two
beams were combined on a beam-splitter and focused on the photodetector using a
3 cm focal length lens. Before reaching the beam splitter, each beam passed
through a variable attenuator formed by a half-wave plate and a polarizing
cube.
The maximum power of each beam at the photodiode was 2 mW. The photodetector
was
a commercial InGaAs Schottky diode with a diameter of 25 micrometers. The
conversion gain was  5V/W.  With an intensity of 0.1 mW per beam, the beat
signal at the output of the photodetector was  -78 dBm, limited by the not
perfect matching of the two beams at the detector.

The beat signal from photodetector PD2 was observed with an rf spectrum
analyzer.
The resolution bandwidth was 10 Hz. The peak frequency could be determined with
a sensitivity of 1 Hz.  After proper thermal stabilization of the apparatus,
the
fluctuations of the values obtained for the frequency of the beat signal were
reduced to less than 1 Hz.

Data were taken by measuring the frequency of the beat note with the spectrum
analyzer while varying the intensity of the laser beams between 0.1 mW and 2 mW
using the variable attenuators.
No dependence of the beat signal frequency on the light intensity was observed
at the accuracy level set by the spectrum analyzer nor any change of the signal
shape.

It is worth stressing that the sensitivity of this method is extremely high. A
fractional change of the frequency of 10$^{-14}$ would have been detected in
this
experiment. In fact, using essentially the same scheme illustrated here, the
accuracy can still be improved by several orders of magnitude by electronical
down-conversion of the beat note and FFT analysis. Also, a higher laser power
and better mode-matching on the photodiode would increase the significancy of
the results.

\section{Discussion}

The null result of this experiment can be translated into an upper bound
for the nonlinearity parameter $~\lambda~$. Assuming ideal conditions
in which all the photons are in a single mode, at the maximum laser
power of 2 mW one obtains a photon number $n \approx 10^{10}$. Using Eq.
(\ref{v2})
and considering the minimum frequency change
$~\frac {\delta \omega }{\omega }\approx 10^{-14}~$ detectable in
this experiment, one obtains $~\lambda \le 10^{-17}~$.
This value is obtained under the ideal conditions mentioned above
and assuming that the frequency of a mode depends only on
the number of photons in that mode. A different bound for
the parameter $~\lambda~$ is obtained, of course, if a different
model is assumed in the interpretation of the experimental result.

Several improvements of the method proposed here are possible. A possibility
already mentioned above is to increase the sensitivity of the apparatus
by electronical down-conversion of the heterodyne signal and FFT
analysis.  Since the
significancy of the results increases with increasing power, the use of a
higher laser power would be desirable. For high laser power, however, a
different photodetector should be used. Such detectors are tipically slow and
this would force to work at a smaller frequency difference. Reduction of the
noise would then require  a long integration time.
As suggested in \cite{sol1}, other types of experiments can be envisaged in
order to test the possibility of describing electrodynamics using
q-oscillators. An accurate test could be performed using interferometric
methods. Another possibility is to study the dependence of the time
correlation function of the q-oscillator on the intensity.

\section*{Acknowledgements}
One of us (V.I.M.) would like to acknowledge the University of
Naples and I.N.F.N., Section of Naples,  for kind hospitality.
G.M.T. acknowledges useful discussions with M. Prevedelli about
the experiment.


\begin{thebibliography}{99}

\bibitem{bie}
L. C. Biedenharn, J. Phys. A {\bf 22}, L873 (1989).

\bibitem{mc}
A. J. Macfarlane, J. Phys. A {\bf 22}, 4581 (1989).

\bibitem{jimbo}
M. Jimbo, Int. J. Mod. Phys. A {\bf 4}, 3759 (1989).

\bibitem{smirnov}
R. N. Alvarez, D. Bonatsos, and Yu. F. Smirnov, Phys. Rev. A {\bf 50},
1088 (1994).

\bibitem{mendes}
V. I. Man'ko and R. V. Mendes, Phys. Lett. A {\bf 180}, 39 (1993).

\bibitem{shabanov}
S. V. Shabanov, {\it Quantum and Classical Mechanics and q-deformed Systems},
Preprint BUTP 92/24 (1992).

\bibitem{vitiello}
E. Celeghini, M. Rasetti, and G. Vitiello, Phys. Rev. Lett. {\bf 66},
2056 (1991).

\bibitem{palak}
J. Lukierski, A. Nowicki, and H. Ruegg, Phys. Lett. B {\bf 293}, 344 (1992).

\bibitem{kostin}
M. D. Kostin, J. Chem. Phys. {\bf 57}, 3589 (1973).

\bibitem{weinberg}
S. Weinberg, Phys. Rev. Lett. {\bf 62}, 485 (1989).

\bibitem{doebner}
H.-D. Doebner and G. A. Goldin, Phys. Lett. A {\bf 162}, 397 (1992).

\bibitem{mizrahi}
V. V. Dodonov and S. S. Mizrahi, Ann. Phys. {\bf 237}, 226 (1995).

\bibitem{su}
G. Su and M. Ge, Phys. Lett. A{ \bf 173}, 17 (1993).ÿ

\bibitem{tuszy}
J. A. Tuszynski, J. L. Rubin, J.Meyer and M. Kibler,
Phys. Lett. A {\bf 175}, 173 (1993).

\bibitem{sol1}
V. I. Man'ko, G. Marmo, S. Solimeno, and F. Zaccaria, Int. J. Mod.
Phys. A {\bf 8}, 3577 (1993).

\bibitem{marmo}
V. I. Man'ko, G. Marmo, and F. Zaccaria, Phys. Lett. A {\bf 191}, 13
(1994).

\bibitem{sol2}
V. I. Man'ko, G. Marmo, S. Solimeno, and F. Zaccaria, Phys. Lett. A
{\bf 176}, 173 (1993); "Q--nonlinearity
of Electromagnetic Field and Deformed Planck Distribution,"
in: {\it Technical Digests of} EQEC'93-EQUAP'93, Firenze, September
10-13, 1993, eds. P. De Natale, R. Meucci, and S. Pelli, vol. {\bf 2}
(1993).

\bibitem{tino}
G.M. Tino, Physica Scripta, T51, 58 (1994).

\bibitem{santarelli}
G. Santarelli, A. Clairon, S.N. Lea, and G.M. Tino, Opt. Commun. 104, 339
(1994).

\end{thebibliography}
\end{document}